\documentclass[preprintnumbers,amsmath,amssymb]{revtex4}

 \usepackage{remreset}

\usepackage[latin1]{inputenc}

\usepackage{graphicx}
\usepackage[bf]{caption2}

\newcommand{\be}{\begin{equation}}
\newcommand{\ee}{\end{equation}}
\newcommand{\bea}{\vspace{0.25cm}\begin{eqnarray}}
\newcommand{\eea}{\end{eqnarray}}

\def\PLA{{Phys. Lett.}  A }
\def\PLB{{Phys. Lett.}  B }
\def\PRL{{Phys. Rev. Lett.} }
\def\PRA{{Phys. Rev.} A }

\def\PRD{{Phys. Rev.} D }

\newlength{\defbaselineskip}
\begin{document}

\title{Cosmology and entanglement}

\author{Marco Genovese}
 \email{m.genovese@inrim.it}

\affiliation{%
I.N.RI.M -- Istituto Nazionale di Ricerca Metrologica\\
Strada delle Cacce, 91 10135 Turin (Italy)
}%
\homepage{http://www.inrim.it/~genovese/marco.html}
\date{\today}

\begin{abstract}
In this paper we present the problem of quantum to classical
transition of quantum fluctuations during inflation and in
particular the question of evolution of entanglement. After a
general introduction, three specific very recent works are discussed
in some more detail drawing some conclusion about the present status
of these researches.

\end{abstract}

\maketitle \vskip 0.5cm Keywords: { entanglement,primordial
fluctuations, inflationary cosmology, decoherence}

\section{Introduction}
 In the last years cosmology rapidly developed thanks to
 observational inputs about far supernovas, large scale structures and cosmic background radiation \cite{physrepc}.

A new paradigm has emerged where big bang model has been
complemented by dark matter, accelerated expansion, inflation
\cite{physrepc}, the so called LCDM (inflationary Cold Dark Matter
model plus cosmological constant).

This new paradigm presents an interesting observational support,
even if the available data would be probably considered at most as
presumptive evidences in other fields, and consistency with
numerical simulation on structure generation \cite{physrepc,str}.

Nevertheless, this paradigm could be by far much weaker than usually
assumed in cosmologists community since most of the very bases of it
lack of a demonstrated physical theory supporting them, for example
dark matter is not yet identified (and also alternatives related to
a modified gravity exist \cite{physrepc}), dark energy related to
accelerated expansion is still rather "obscure" \cite{asl}, etc.

Even the first element of this paradigm, inflation \cite{inf}, not
only still lacks of a confirmed physical theory describing it, but
also has various problems to be solved at its very basis. The most
important is probably the transition from quantum to classical
fluctuations.

In little more detail the primordial spectrum of perturbations,
which represent the seed for the developing of structures in the
universe, is usually assumed to be described by a classical
distribution function. However, they are supposed to derive from
generation of short wave length quantum fluctuations of the inflaton
field \cite{1,2,3}. Thus, the transition from a quantum to a
classical distribution must be justified. In particular, as
emphasized recently \cite{1}, this transition must be considered
keeping into account the entanglement of vacuum, indeed quantum
fluctuations cannot be regarded as stochastic fluctuations as long
as the system is entangled, in principle even violations of Bell
inequalities could be measured in this situation \cite{5}. On the
other hand entanglement among modes is an unavoidable consequence of
evolution equation \cite{5}. This requires to specifically analyze
the evolution of some entanglement measure and not only of other
parameters used to quantify classicality, as previously done
\cite{gau} (for further criticisms to these previous works see also
\cite{sud}). As a further argument on the relevance of entanglement,
not mentioned previously as fa as we now, we would like also to hint
at the fact that quantum field theory vacuum is expected to be in an
entangled state leading to violation of some Bell inequality
\cite{wer}.

Here we will review the most recent progresses about this specific
problem \cite{1,2,3,4} and present some personal consideration. We
will try to avoid to enter technical details, trying to emphasize
main assumptions and limits of the presented studies from a quantum
mechanical point of view.

The outline of the paper is the following, in section II we
introduce the concept of entanglement and its quantification, in
section III we epitomize the ideas about quantum to classical
systems transition. Then some recent models for quantum to classical
perturbations transition are summarized: in section IV Nambu's
model, in section V Campo-Parentani one and in section VI De
Un\'anue-- Sudarsky model. Finally, in section VII we draw some
conclusion.

\section{Something about entanglement}

Entanglement was defined by Schr\"odinger "the characteristic trait
of quantum mechanics". A system of many particles is defined as
entangled when the global wave function cannot be factorized in
single particle wave functions \cite{mg}. This means that the system
must be considered as a whole and a measurement on one subsystem
instantaneously influences the following measurements on other
subsystems (it can be rigorously demonstrated that this does not
allow superluminal transmission of information \cite{mg}).

This property has been at the very basis of many of the discussions
about foundations of quantum mechanics (QM) and now it is seen as a
resource for developing quantum technologies \cite{NC} or a tool for
studying phase transitions (see \cite{p} and ref.s therein). Thus a
huge literature exits on this field. For our purposes we limit to
consider the case of entanglement of continuous variables states
\cite{mt}.

Thus, we consider n bosonic modes described by annihilation
operators $a_n$  satisfying usual commutation relations. Then one
defines the phase space variables: \bea q_n = (a_n +
a_n^{\dagger})/2
\\p_n= (a_n - a_n^{\dagger})/(2 i) \cr
S=(q_1,p_1,...,q_n,p_n) \eea and a covariance matrix: \be V_{ij} =
1/2 [\langle \{S_i,S_j \} \rangle -\langle S_i \rangle \langle S_j
\rangle ] \ee  $\{.,. \}$ being the anticommutator.

Defined the symplectic matrix \be \Omega = \left(
\begin{array}{lllll}
 0 & 1 & ...& 0 & 0 \\
 -1 & 0 & ..&0 & 0 \\
  & &...& & \\
 0 & 0 &...& 0 & 1 \\
 0 & 0 &...& -1 & 0
\end{array}
\right) \ee for physical systems one has \be V + { i \over 2} \Omega
\geq 0 \ee
 On the other hand a generic bipartite mixed state is separable if its density matrix
$\rho$ is of the form \be  \rho = \sum_i (\rho^A_i \bigotimes
\rho^B_i) w_i \label{sep} \ee where $\sum_i w_i = 1$, $w_i > 0$ and
where $\rho^A_i$ $\rho^B_i$ are the density matrices of subsystems A
and B. Every non-separable state is called entangled.

A general criterium for separability does not exist yet. Anyway, for
a gaussian state (i.e. a state whose Wigner function is Gaussian) it
is the Simon's one \cite{sim}, \be \Lambda V \Lambda^T + { i \over
2} \Omega \geq 0 \label{sim}\ee where we have defined $\Lambda =
diag (1,-1,...,1,-1)$.

  The matrix  $\Lambda V \Lambda^T$ can be
diagonalized by an appropriate symplectic transformation, e.g to
$diag(\lambda_+,\lambda_+,\lambda_-,\lambda_-)$ for a two modes
gaussian state.

Considering two modes gaussian states, in terms of its symplectic
eigenvalues the condition \ref{sim} reads \be\lambda_- \geq 1/2
\label{sim2}\ee

 On the other hand various measures of entanglement have
been proposed \cite{z,mt}. One has to mention that the hierarchy of
entanglement is entanglement-measure dependent \cite{z,ad} and that
all of them present both advantages and disadvantages \cite{z}.
Among them one that presents various useful properties is
negativity, defined as (denoting $||{A}||_{tr} \equiv
Tr[\sqrt(A^{\dagger }A)]$) \be N(\rho) \equiv 1/2 ( ||\rho
^{T_A}||_{tr}-1) \ee where $T_A$ denotes the partial transposition
of subsystem $A$. Often also the related logarithmic negativity
$E_N\equiv \ln  ||\rho ^{T_A}||_{tr}$ is used. For a bipartite
gaussian system in terms of symplectic eigenvalues one has: \be E_N
= - min [log_2(2 \lambda_-),0] \ee that is a decreasing function of
the smallest partially transposed symplectic eigenvalue $\lambda_-$,
that therefore completely qualifies and quantifies the entanglement
of a two-mode gaussian state.

\section{Hints at the decoherence problem}

The problem we are considering is the transition from quantum states
to classical ones during universe evolution, i.e. the developing of
decoherence: this is connected to a long debated problem at the
foundations of quantum mechanics pertaining the transition from a
microscopic quantum world to a macroscopic classical one (the so
called macro-objectivation problem), that indeed did not found yet a
conclusive solution \cite{aul,ghi,sch}.

This problem derives from the fact that Schr\"odinger equation is
linear and thus requires that a macroscopic system interacting with
a superposition state  originates an entangled state including the
macroscopic system as well.

In order to introduce, in a very synthetic way, the question let us
consider a macroscopic system described by the state $| \Phi _0 >$
that interacts with the (microscopic) states $ \vert \chi_1 \rangle
$ and $ \vert \chi_2 \rangle $. The interaction, lasting a time
interval $\Delta t$, is described by a linear evolution operator
$U(\Delta t)$ and leads to

\be \vert \Phi_0 \rangle \vert \chi_1 \rangle \rightarrow U(\Delta
t) [\vert \Phi_0 \rangle \vert \chi_1 \rangle ] = \vert \Phi_1
\rangle \vert \chi_1 \rangle \ee

and \be \vert \Phi_0 \rangle \vert \chi_2 \rangle \rightarrow \vert
\Phi_2 \rangle  \vert \chi_2 \rangle \ee where the states $\vert
\Phi_1 \rangle$ and $\vert \Phi_2 \rangle$ represent the state of
the macroscopic system after interaction.

If $| \Phi _0 >$ interacts with the superposition state \be a \vert
\chi_1 \rangle + b \vert \chi_2 \rangle \ee because of linearity of
the evolution equation, one has \be \vert \Phi_0 \rangle [a \vert
\chi_1 \rangle + b \vert \chi_2 \rangle ] \rightarrow[a \vert \Phi_1
\rangle  \vert \chi_1 \rangle +b \vert \Phi_2 \rangle  \vert \chi_2
\rangle ] \label{vns} \ee that is an entangled state involving the
macroscopic system as well.

Of course at macroscopic level one does not observe superpositions
of different macroscopic states, thus one expects that at some point
the wave function "collapses" to a defined state, i.e. only one
state in the superposition survives: in the previous example the
classical system is either in the situation described by $\vert
\Phi_1 \rangle$, with probability $|a|^2$, or in the one described
by the state $\vert \Phi_2 \rangle $, with probability $|b|^2$.
However, this request must be justified more precisely.

Different answers to macro-objectivation problem have emerged. A
first possibility is to split the world into a macroscopic one
following classical mechanics and a microscopic one following
quantum mechanics (substantially the one adopted by the Copenaghen
school). However this solution, even if perfectly useful for all
practical calculation of quantum processes, does not really solve
the problem since it does not answer to the fundamental question of
how and when the quantum world becomes classical.

 Therefore, various different ideas have been considered for
 explaining/understanding decoherence at macroscopic level, without reaching for any of them
a general consensus in the physicists community \cite{ghi}. Among
them (without any pretence to be exhaustive) one can mention: the
many universes models \cite{everett} (each state is realized in
non-communicating universes) , modal interpretations (one tries to
attribute definite values at certain sets of observables)
\cite{modal1,modal2,modal3}, decoherence and quantum histories
schemes  \cite{z1,gr,zurek,GMH} (coherence is preserved and
classical level, but unobservable), transactional interpretation
\cite{cram}, 'informational' interpretation (QM concerns
"information" not reality) \cite{zeiinf,fuchs}, dynamical reduction
models (where a non-linear modification of Schr\"odinger equation is
introduced) \cite{GRW,pearle,mil}, reduction by consciousness (wave
function collapse happens at observer level) \cite{wigner} and many
others (see for example \cite{aul,rod,par,bub,HW,fpp} and Ref.s
therein).

Let us just epitomize  a little more in detail the schemes that can
have some interest in our discussion.

In general the considerations on quantum to classical transition
\cite{gau,1,2,3} in cosmology adopt the idea that quantum mechanics
is valid without any change at every scale. Here, the evolution is
always unitary and the interacting macroscopic states remain
entangled, but the interference cannot be observed at macroscopic
level since the macroscopic states rapidly become orthogonal. In
particular when environmental degrees of freedom are included the
density matrix of the system under analysis is obtained after having
traced off the environment degrees of freedom: since environment
states rapidly become orthogonal one is left with a statistical
mixture for the system. This is the "decoherence" framework
\cite{z1,gr,zurek,GMH}. Various model and physical examples have
been considered for demonstrating this scenario. Of course, it has
the advantage that quantum mechanics is assumed to be valid as it is
without any change, wave function collapse is simply eliminated from
the discussion as "not needed". Somehow this scheme may be
reconnected to "many worlds" one. Nevertheless, many physicists
studying foundations of quantum mechanics are not satisfied by this
model \cite{gr,adl,b}. The main objections are that:

- linear superpositions of macroscopic states occur anyway, to
replace them with statistical mixtures is only a trick "valid for
most of practical purposes".

- in quantum mechanics the correspondence between statistical
ensembles and statistical operators is infinitely many to one.
Different statistical mixtures (also containing macroscopic states
superpositions) correspond to the same statistical operator.

Thus, decoherence scheme, as admitted by the same proposers
\cite{z}, remain not clearly justified.

 On the other hand, macro-objectivation problem simply does not exist in Hidden Variable Models (HVM)
\cite{mg,thooft} since in this case the specification of the state
by using state vectors is insufficient, there are further parameters
(the hidden variables), which we ignore, for characterizing the
physical situation: in this case the physical system is always in a
well specified state univocally determined by the value of the
hidden variables. One must mention that if Bell inequality tests
strongly disfavour \footnote{some doubts remaining due to unsolved
detection loophole (or in the case of ions being the measurements
not separated) \cite{mg}.} local HVM \cite{mg}, they do not concern
non-local HVM, as de Broglie-Bohm or Nelson's model, or high energy
(Planck) scale HVM \cite{mg}.

As an example of non-local HVM, in dBB framework the hidden variable
is the position of the particle that evolves with Hamilton-Jacobi
equations including also a "quantum potential" related to the wave
function ($\Psi =R(x,t)\cdot exp[iS(x,t)/ \hbar]$, evolving with
standard quantum equation):
\bigskip
\begin{equation}
\frac{\partial S}{\partial t}+\frac{(\nabla S)^{2}}{2m}+V+Q=0
\end{equation}
 The presence of the quantum potential $Q=-\frac{\hbar ^{2}}{2m}\frac{\Delta
 R}{R}$ implies that the trajectory is "instantaneously" affected by any
change in the system where the wave function is non-zero (the
non-locality). In principle in dBB the evolution is the same as in
standard quantum mechanics \footnote{see Ref. \cite{nosdBB} and
ref.s therein for some alternative point of view.}, but the single
particle is never in a superposition, the QM probabilistic structure
derives from the ignorance of the hidden variables values (i.e. the
positions), values that even in principle cannot be determined
(otherwise superluminal signalling would be possible).
Generalization to relativistic case is not trivial (e.g. see
\cite{ghi} and ref.s therein).

Rather different are HVM at large scales. Here the idea is that at
large energy (e.g. Planck) scales physical systems (including
gravity) are described by a deterministic theory (non-unitary
evolution), but at smaller scales we have loss of information (for
dissipation): "quantum states" are equivalence classes of the
deterministic states, the loss information are the hidden variables.
The evolution of these equivalence classes of states becomes
unitary. Present tests of Bell inequalities do not exclude these
models since not pertaining the "correct" degrees of freedom. Of
course in these models
\cite{thooft,blas,bru,bla2,bla3,thooft2,thooft3,bm03,elz,kato,wet}
(for the moment far from reaching a definite theory) "decoherence"
at large energy scale should be properly treated using the models
themselves \footnote{A similar situation happens also for
"pre-quantum" theory of Adler \cite{adl2}.}.

Finally, in dynamical reduction models \cite{GRW,pearle,mil} the
wave function collapse is introduced by modifying the quantum
evolution equation with a non-linear term (the collapse can
eventually be related to Quantum Gravity \cite{pen}, see also
Ref.\cite{har} for cosmological implications of modifications of QM
related to QG).

For example in one of the seminal models \cite{GRW}  the wave
function suddenly randomly collapses according to

\bea \Psi(x_1,...,x_N) j(x-x_i) / R \label{qc} \\
 j(x-x_i) = A \exp [-(x-x_i)^2/ (2 a)^2] \label{j} \,\,\,\,\,\,\, |R(x)|^2 = \int dx_1 ...  dx_N |\Psi(x_1,...,x_N) j(x-x_i)|^2
\eea where  $x_i$ is the specific coordinate of the $i$th particle
of the system (the one undergoing the collapse). The probability of
the collapse is given, for each particle, by $1/\tau$, where $\tau$
can be fixed to be $\approx 10^{15} s \approx 10^{8}$ years; the
constant $a$ should be $a \approx 10^{-7}$ m and the collapse center
$x$ is randomly chosen with probability distribution $|R(x)|^2 $.

These collapse models usually comport small (not observable with
present experiments) violation of energy conservation in
non-relativistic versions, but these become difficult to tame in
extensions to relativistic case. Obviously in these models
decoherence should be kept into account by considering the modified
evolution equation and could take a long time before happening.

 As mentioned, almost all the studies about transition to classical perturbations are
inside the decoherence scheme, however one should keep in mind that
alternative schemes as dynamical reduction models or HVM could give
different scenarios. As far as we know this point was considered
only in a couple of papers of Sudarsky et al. \cite{4,sud} where a
specific wave function collapse model was studied in relation to the
generation of seeds of cosmic structures (presented in section VI).

\section{Nambu's approach}
As a first example of a recent study on the transition from a
quantum to a classical spectrum, we consider Ref.\cite{1}.

He considers a real massless scalar field $\phi$ in a de Sitter
universe (i.e. with no ordinary matter content but with a positive
cosmological constant), corresponding to a metric and a Lagrangian:
\bea ds^2= a(\eta)^2 (-d \eta^2 + dx^2), \, \, \,\, \, \, a= {- 1
\over H \eta}  \cr - \infty < \eta < 0, \, \, \,\, \, \, L= \int d^3
x {1 \over 2} \sqrt{-g} \left(-{1 \over 2} g^{\mu \nu}
\partial_{\mu} \phi \partial_{\nu} \phi \right )\label{fi}\eea
where $\eta \equiv \int {dt \over a(t)}$ is conformal time and H the
Hubble constant.

In terms of the conformally rescaled variable $q= a \phi$ one has
the usual equation of motion \be q"-{a" \over a} q - \partial _i^2 q
= 0 \label{em}\ee

A discrete unidimensional lattice model of the scalar field, that
regularizes ultraviolet divergences, is adopted. The quantized
canonical variables are represented as: \bea q_j= {1/\sqrt{N}}
\sum_{K=0}^{N-1} (f_k a_k + f^*_k a^{\dagger}_{N-k})e^{i \theta_k j}
 \cr p_j= {-i/\sqrt{N}} \sum_{K=0}^{N-1} (g_k a_k - g^*_k
a^{\dagger}_{N-k})e^{i \theta_k j} \,\,\,\,\,\,\,\,\,\,\,\,
\theta_k={2 \pi k \over N}\label{q}\eea with the usual commutation
relations $[a_k,a^{\dagger}_j]=\delta_{k,j}, ...$.

Then the Bunch-Davis \cite{BD} vacuum (i.e. the one with only
positive frequencies in asymptotic past) is assumed: \be f_k={1
\over 2
\omega_k} (1 + {1 \over i \omega_k \eta} ) e^{-i \omega_k \eta}, \\
g_k=\sqrt{\omega_k \over 2} e^{-i \omega_k \eta}\label{BDv} \ee

In order to reduce to a bipartite entanglement the next step is
introducing block variables pertaining two spatial regions $A,B$
(containing $n$ lattice sites):\be q_{A (B)}= {1 \over \sqrt{n}
}\sum_{j\in A (B)} q_j \ee

Finally, the covariance matrix is evaluated and the separability
criterium \ref{sim2} is applied. The fact to limit the calculation
to two regions implies a non-unitary evolution.

The numerical simulation (on a 100 sites lattice) shows that the
logarithmic negativity, initially different from 0 pointing out an
entanglement between regions A and B (when the initial distance
between the borders of the two regions is zero), always goes to zero
(separable states) when $\eta$ grows (reaching zero essentially when
the the physical size of each region exceeds the Hubble horizon
length).

For the sake of completeness, we can mention that also other
"classicality" tests are considered, as positivity of P-function and
identity of P and Wigner functions (for a discussion on the
hierarchy of classicality measures for gaussian states see
Ref.\cite{nos}).

In summary, inside Nambu's model a transition from quantum to
classical perturbations happens. Nevertheless, this result, albeit
interesting, relies on certain approximations. First of all, as the
same author mentions, he considers the Bunch-Davis vacuum. Thus the
results concerns this specific choice; in particular what would
happen for a non--gaussian initial state remains unknown. The
analysis of Lesgourgues et al. \cite{les} shows that for an initial
non-gaussian state one has a transition to a semiclassical behaviour
in the sense that the Wigner function, albeit remaining non positive
in some regions, becomes concentrated near classical trajectories.
This can also be rephrased by saying that the quantum field becomes
a highly squeezed state, for which the non-commutativity between
conjugated variables can be neglected (for analogous conclusion for
gaussian state see also \cite{gau}). However, this analysis is
limited since it does not consider entanglement and, in absence of
an entanglement criterium for non-gaussian states, this cannot be
done even in principle.

Besides this point, other weak points are present in Nambu's model
\cite{1}. We would like just to mention:

 - the non-unitary evolution due to limiting the analysis to regions A and B must be considered
 carefully. Entanglement evolution of systems in a bath can presents
 non-trivial aspects as a "revival" of entanglement after this had disappeared \cite{rev}.

 - the simplification of a monodimensional model on lattice.

\section{Campo-Parentani model}

First of all Campo and Parentani \cite{2} mention the problems of
previous approaches (substantially those in ref. \cite{gau}). The
main identified problems are

- the use of a master equation, since it is neither clear how to
properly define it (due to renormalization) nor how to trace over
"unobservables" degrees of freedom.

- the introduction of an "environment".

 They argue to solve these problems by using Green function method, since, on the one hand,
this overcomes the use of master equations and, on the other hand,
allows defining a clear intrinsic coarse graining procedure. Respect
to Ref.\cite{1} less emphasis on the necessity of strictly
considering the entanglement is given, even if then they state that
"the quantum to classical-transition thus occurs when this [between
modes of opposite wave vectors] entanglement is lost".

 The physical situation they consider \cite{2,3} is
exactly the one presented in the previous paragraph: study of the
evolution of entanglement starting from the Bunch-Davis vacuum
\cite{BD}. Again the covariance matrix and its evolution are
evaluated and a coarse graining is defined (through a truncation of
the hierarchy of Green functions). The Von Neumann entropy, $S=-
tr[\rho ln(\rho)]$ (valuing 0 for a pure state), associated with the
coarse graining gaussian two modes reduced density matrices is used
as measure of entanglement (in order to motivate this choice, the
authors show that $S$ is gauge invariant and independent of
renormalization), being $S(n)\simeq ln(n)$ (n being the average
particles number) at the threshold of separability deriving form
Simon's criterium. On the other hand it is shown that the criterium
of separability depends on the choice of canonical variables.

 They focus on curvature perturbations $\zeta$
(defined as \cite{cur} $g_{ij}\simeq a^2(\eta) \delta_{ij} e^{2
\zeta}$) relating the growth in entropy to the covariance matrix of
$\zeta$. A little more in detail in linear approximation the
evolution of $\zeta$ is determined by the Lagrangian:
 \be L = {1 \over 8 \pi G} \int d^3 x a^3 {-\partial H/\partial t \over H^2} \left[(\partial{\zeta}/\partial
 t)^2-(\nabla \zeta)^2/a^2 \right]
 \ee
 The free vacuum is assumed to be the Bunch-Davis one:
\be (i \partial _{\tau} - q) \left(a \sqrt {-\partial H/\partial t
\over H^2} \zeta_q \right)\rightarrow 0, \,\,\,\,\,\, {q \over a H }
\rightarrow \infty \ee where $\zeta_q$ is the Fourier mode with
comoving wave vector $q$.

Then the covariance matrix ($\pi_{-q}$  being the momentum
conjugate) \bea C = {1\over 2} Tr[\rho \{V,V^{\dagger}\}]
\,\,\,\,\,\,\, V = \left( \begin{array}{l}
 \zeta_q \\ \pi_{-q}
\end{array} \right) \,\,\,\,\,\,\,\,\,\,
\eea is evaluated by the Green functions of reduced density matrices
(that are gaussian when  using a coarse graining truncating Green
function hierarchy at order 2) $G(t,t',q) \delta^3(q-q') \equiv
{1\over 2} Tr[\rho^{red}_{q,-q} \{\zeta_q(t),\zeta_{-q'}(t') \}]$,
being $C_{1,1} = G(t,t,q), ....$ \cite{3}. The entropy of
$\rho^{red}_{q,-q}$ ir related to the determinant of the covariance
matrix by $S=ln[det (C)]$.

After having derived the equation for the evolution of $S$, this is
applied to different situations \cite{3}.

One considered case is multifield inflation, in particular the two
field model: \bea L= \int d^3 x  \sqrt{-g} \left [{R \over 16 \pi
G}-{1 \over 2} g^{\mu \nu} (
\partial_{\mu} \phi \partial_{\nu} \phi + e^{2 b(\phi)} \partial_{\mu} \chi \partial_{\nu} \chi) -V(\phi,\chi)  \right ]\label{fim}\eea

Defined the field $\sigma$
\be
\partial_t \sigma^2 =\partial_t \phi^2 + e^{2 b}\partial_t \chi^2
\ee with ($\Phi$ being the gravitational potential) \be ds^2 = -(1+
2 \Phi) dt^2 + a^2 (1-2 \Phi) \delta_{ij} dx^i dx^j \label{mp}\ee
one has \be \zeta = \Phi + {H \over \partial_t \sigma} \delta \sigma
\ee

 Another analyzed case is a single inflaton field model plus a
matter field $\sigma$ (in a fundamental representation of $O(N)$),
whose quadratic part of the Lagrangian describing the free evolution
is:\be L= {1 \over 2} \sum_{n=1}^N \int d^3 x a^3 \left[ {d \sigma_n
\over dt}^2-{1 \over a^2} (\nabla \sigma_n)^2 \right]\ee for a
minimal coupling to gravity.

 Without entering into the details of the calculation, the
conclusions are that in the first case entropy grows at a high rate,
both in presence of an anomalous kinetic term ($b(\phi)\neq 0$) or
in its absence ($b(\phi)= 0$), reaching  at the end of inflation a
value largely above the separability threshold $S_T= 4 N_{end}$
($N_{end}$ being the number of e-folds from horizon exit to the end
of inflation). Besides the authors argue that this result can be
extended also to inflationary models with many fields at least when
the kinetic term is in the canonic form. On the other hand the
single field model does not show evidences of decoherence since von
Neumann entropy of reduced density matrix remains small.

For the sake of completeness, we can also mention that, similarly to
Ref.\cite{1}, other criteria of classicality are considered as well,
as broadness of Wigner function (i.e. the Wigner function is broad
enough to be the Husimi representation of some normalizable density
matrix), existence of a P-representation, etc. showing how these
lead to different thresholds: this induces the authors to state that
there is no intrinsic threshold of decoherence at which
quantum--to--classical transition occurs.

For what concerns the generality of these results, one has to
mention that the first two considerations about Nambu's model (on
gaussian vacuum and coarse graining) completely keep their validity
here as well.

\section{De Un\'anue--Sudarsky model}

In Ref. \cite{4} De Un\'anue--D. Sudarsky consider a specific wave
function collapse model, improving a precedent scheme \cite{sud}.

The starting point is always to consider inflation with scalar field
with a Lagrangian like Eq. \ref{fi},  a metric like Eq. \ref{mp} and
a Bunch-Davis vacuum Eq. \ref{BDv}.

Then a collapse is introduced: before the collapse there are no
metric perturbations, it is only after the collapse that the
gravitational perturbations appear, i.e the collapse of each mode
represents the onset of inhomogeneity and anisotropy at the scale
represented by the mode.

Of course introducing the collapse solves completely the problem of
quantum to classical transition: after the collapse the Universe is
in a defined single state (and of course entanglement has
disappeared).

The specific collapse a certain time in a state $|\Omega \rangle$ is
such that one finds for the expectation values of the variable
defined in Eq. \ref{q} \be \langle q_k^R \rangle _\Omega = x_k
\Lambda_k \cos \Theta_k, \,\,\,\, \langle p_k^R \rangle _\Omega =
x_k \Lambda_k  k \sin\Theta_k \ee where $x_k$ is a random variable,
$\Lambda_k$ is the major semi-axis of the ellipse corresponding to
the boundary of the region in phase space where the Wigner function
is lager than 1/2 its maximum value, $\Theta_k$ is the angle between
that axis and the $f_k^R$ axis.

From this expressions one calculates the fluctuations $ \langle
\delta \phi_k \rangle$. The perturbation spectrum is then compared
with what needed by present numerical simulations of evolution of
structures. The authors remark that this analysis is just
preliminary and more refined studies are needed. Their results show
how the power spectrum is sensible to the choice of the collapse
model, being in agreement with what expected for certain choices and
clearly in disagreement for others: thus  different models could be
selected by comparison with observational data.

\section{Concluding Remarks}

In conclusion our opinion is that a clear  demonstration of the
transition to classical from quantum perturbations at the end of
inflation is still lacking.

Since, for the considerations expressed in the introduction, one
should unavoidably consider entanglement of vacuum, we have limited
a specific analysis to the most recent papers where this was done,
even if most of the considerations expressed about these models also
pertain previous studies \cite{gau}. A first point to be mentioned
is that the results are model dependent. Then, all of them consider
gaussian states, but this is a limiting hypothesis that has no
motivation.

These two problems already sustain our conclusion. But the situation
is even worse than that, since in almost all the studies about this
phenomenon \cite{1,2,3,gau} a decoherence model for
quantum-classical transition was assumed, but in principle
completely different results could derive if other models would be
applied (as dynamical reduction one or HVM). A first attempt in this
sense \cite{5} was presented in the last section, where a scheme
considering a specific wave function collapse model was presented.
However, of course this scheme is just a first step of this possible
analysis and it is pertaining a very specific framework.

Thus, in summary, in our opinion, the transition to classical from
quantum perturbations remains (and probably will remain for a long
time up to when we will not dispose of a well proven theory at the
scale of inflation) one of the unsolved problem of present
cosmological models.

  \vskip 0.5cm
\textbf{Acknowledgements}: This work has been supported by  Piedmont
Region (project E14).

\end{document}